\documentclass[english,aps,manuscript]{revtex4-1}
\usepackage[T1]{fontenc}
\usepackage[latin9]{inputenc}
\usepackage[active]{srcltx}
\usepackage{amsmath}
\usepackage{esint}

\makeatletter

\DeclareFontEncoding{LGR}{}{}
\DeclareTextSymbol{\~}{LGR}{126}
\newcommand{\lyxmathsym}[1]{\ifmmode\begingroup\def\b@ld{bold}
  \text{\ifx\math@version\b@ld\bfseries\fi#1}\endgroup\else#1\fi}

\@ifundefined{textcolor}{}
{%
 \definecolor{BLACK}{gray}{0}
 \definecolor{WHITE}{gray}{1}
 \definecolor{RED}{rgb}{1,0,0}
 \definecolor{GREEN}{rgb}{0,1,0}
 \definecolor{BLUE}{rgb}{0,0,1}
 \definecolor{CYAN}{cmyk}{1,0,0,0}
 \definecolor{MAGENTA}{cmyk}{0,1,0,0}
 \definecolor{YELLOW}{cmyk}{0,0,1,0}
 }

\makeatother

\usepackage{babel}
\begin{document}

\title{Geometric Thermodynamics of Schwarzschild-AdS black hole with a Cosmological
Constant as State Variable}

\author{Alexis Larrañaga}

\address{National Astronomical Observatory. National University of Colombia.}

\author{Alejandro C\'ardenas}

\address{Physics Department National University of Colombia.}

\begin{abstract}
The thermodynamics of the Schwarzschild-AdS black hole is reformulated
within the context of the recently developed formalism of geometrothermodynamics
(GTD). Different choices of the metric in the equilibrium states manifold
are used in order to reproduce the Hawking-Page phase transition as
a divergence of the thermodynamical curvature scalar. We show that
the enthalpy and total energy representations of GTD does not reproduce
the transition while the entropy representation gives the expected
behavior.


\end{abstract}
\maketitle

\section{Introduction}

The thermodynamics of black holes has been studied extensively since
the work of Hawking \cite{key-1}. The notion of critical behavior
has arisen in several contexts for black holes, ranging from the Hawking-Page
\cite{key-2} phase transition in anti-de-Sitter space and the pioneering
work by Davies \cite{key-3} on the thermodynamics of Kerr-Newman
black holes, to the idea that the extremal limit of various black
hole families might themselves be regarded as genuine critical points
\cite{key-4,key-5,key-6}. In most treatments of black-hole thermodynamics
the cosmological constant, $\Lambda$, is treated as a fixed parameter
(possibly zero) but it has been considered as a dynamical variable
in \cite{lambda1,lambda2} and it has further been suggested that
it is better to consider $\Lambda$ as a thermodynamic variable, \cite{lambda3,lambda4,lambda5,lambda6,quevedo08-1}.
Physically, $\Lambda$ is interpreted as a thermodynamic pressure
in \cite{entalpia1,entalpia2}, consistent with the observation in
\cite{entalpia3,dolan,cvetic} that the conjugate thermodynamic variable
is proportional to a volume. 

Now, the use of geometry in statistical mechanics was pioneered by
Ruppeiner \cite{rup79} and Weinhold \cite{wei1}, who suggested that
the curvature of a metric defined on the space of parameters of a
statistical mechanical theory could provide information about the
phase structure. However, some puzzling anomalies become apparent
when these methods are applied to the study of black hole thermodynamics.
A possible resolution was suggested by Quevedo\textquoteright{}s geometrothermodynamics
(GTD) whose starting point \cite{quev07} was the observation that
standard thermodynamics was invariant with respect to Legendre transformations,
since one expects consistent results whatever starting potential one
takes. The formalism of GTD indicates that phase transitions occur
at those points where the thermodynamic curvature is singular, but the results of Quevedo also  show that the metric structure of the phase manifold determines the type of systems that can be described by a specific thermodynamic metric. For example, an Euclidean structure describes systems with first order phase transitions, whereas a pseudo-Euclidean structure describes systems with second order phase transitions. Actually, there is no complete explanation for this result but it is clear that the phase manifold contains information about thermodynamic systems.

In this paper we apply the GTD formalism to the Schwarzschild-AdS
black hole to investigate the behavior of the thermodynamical curvature.
As is well known, a black hole with a positive cosmological constant
has both a cosmological horizon and an event horizon. These have different
Hawking temperatures associated with them in general, which complicates
any thermodynamical treatment. Therefore we will focus on the case
of a negative cosmological constant, though many of the conclusions
are applicable to the positive $\Lambda$ case. Even more, the negative
$\Lambda$ case is of interest for studies on AdS/CFT correspondence
and the considerations here are likely to be relevant in those studies.
Trying  with different thermodynamical potentials and choices of the metric in the equilibrium states manifold
we will show that the entropy representation of GTD is the one that reproduces the
Hawking-Page phase transition as a divergence in the thermodynamical
curvature scalar and it appears to be a second order transition.

\section{Geometrothermodynamics in Brief}

The formulation of GTD is based on the use of contact geometry as
a framework for thermodynamics. Consider the $(2n+1)$-dimensional
thermodynamic phase space $\mathcal{T}$ coordinatized by the thermodynamic
potential $\Phi$, extensive variables $E^{a}$, and intensive variables
$I^{a}$, with $a=1,...,n$. Let us define on $\mathcal{T}$ a non-degenerate
metric $G=G(Z^{A})$, with $Z^{A}=\{\Phi,E^{a},I^{a}\}$, and the
Gibbs 1-form $\Theta=d\Phi-\delta_{ab}I^{a}dE^{b}$, with $\delta_{ab}={\rm diag}(1,1,...,1)$.
If the condition $\Theta\wedge(d\Theta)^{n}\neq0$ is satisfied, the
set $(\mathcal{T},\Theta,G)$ defines a contact Riemannian manifold.
The Gibbs 1-form is invariant with respect to Legendre transformations,
while the metric $G$ is Legendre invariant if its functional dependence
on $Z^{A}$ does not change under a Legendre transformation. Legendre
invariance guarantees that the geometric properties of $G$ do not
depend on the thermodynamic potential used in its construction. 

The $n$-dimensional subspace $\mathcal{E}\subset\mathcal{T}$ is
called the space of equilibrium thermodynamic states if it is determined
by the smooth mapping 
\begin{eqnarray}
\varphi:\ \mathcal{E} & \longrightarrow & \mathcal{T}\nonumber \\
(E^{a}) & \longmapsto & \left(\Phi,E^{a},I^{a}\right)
\end{eqnarray}
with $\Phi=\Phi(E^{a})$, and the condition $\varphi^{*}(\Theta)=0$
is satisfied, i.e. 
\begin{equation}
d\Phi=\delta_{ab}I^{a}dE^{b}
\end{equation}
 
\begin{equation}
\frac{\partial\Phi}{\partial E^{a}}=\delta_{ab}I^{b}.
\end{equation}
The first of these equations corresponds to the first law of thermodynamics,
whereas the second one is usually known as the condition for thermodynamic
equilibrium (the intensive thermodynamic variables are dual to the
extensive ones). The mapping $\varphi$ defined above implies that
the equation $\Phi=\Phi(E^{a})$ must be explicitly given. This relation
is known as the fundamental equation from which all the equations
of state can be derived. Finally, the second law of thermodynamics
is equivalent to the convexity condition on the thermodynamic potential,
\begin{equation}
\partial^{2}\Phi/\partial E^{a}\partial E^{b}\geq0.
\end{equation}

The thermodynamic potential satisfies the homogeneity condition $\Phi(\lambda E^{a})=\lambda^{\beta}\Phi(E^{a})$
for constant parameters $\lambda$ and $\beta$. Therefore, it satisfies
the Euler's identity, 
\begin{equation}
\beta\Phi(E^{a})=\delta_{ab}I^{b}E^{a},
\end{equation}
and using the first law of thermodynamics, we obtain the Gibbs-Duhem
relation, 
\begin{equation}
(1-\beta)\delta_{ab}I^{a}dE^{b}+\delta_{ab}E^{a}dI^{b}=0.
\end{equation}

We also define a non-degenerate metric structure $g$ on ${\cal E}$,
that is compatible with the metric $G$ on ${\cal T}$. Now we will
formulate the main statement of geometrothermodynamics. A thermodynamic
system is described by a metric $G$ which is called a thermodynamic
metric \cite{quev07} if it is invariant with respect to transformations
which do not modify the contact structure of ${\cal T}$ . In particular,
$G$ must be invariant with respect to Legendre transformations in
order for GTD to be able to describe thermodynamic properties in terms
of geometric concepts in a manner which must be invariant with respect
to changes of the thermodynamic potential. In the language of GTD,
a partial Legendre transformation is written as

\begin{equation}
Z^{A}\rightarrow\tilde{Z}^{A}=\left\{ \tilde{\Phi},\tilde{E}^{a},\tilde{I}^{a}\right\} 
\end{equation}
where

\begin{equation}
\begin{cases}
\Phi & =\tilde{\Phi}-\delta_{kl}\tilde{E}^{k}\tilde{I}^{l}\\
E^{i} & =-\tilde{I}^{i}\\
E^{j} & =\tilde{E}^{i}\\
I^{i} & =\tilde{E}^{i}\\
I^{j} & =\tilde{I}^{j},
\end{cases}
\end{equation}

with $i\cup j$ any disjoint decomposition of the set of indices $\left\{ 1,2,...,n\right\} $
and $k,l=1,...,i$.

As is shown in \cite{quev07}, a Legendre invariant metric $G$ induces
a Legendre invariant metric $g$ on ${\cal E}$ defined by the pullback
$\varphi^{*}$ as $g=\varphi^{*}(G)$. There is a vast number of metrics
on ${\cal T}$ that satisfy the Legendre invariance condition. The
results of Quevedo et al. \cite{quevedo08,quevedo09,quevedo10} show
that phase transitions occur at those points where the thermodynamic
curvature is singular and that the metric structure of the phase manifold
${\cal T}$ determines the type of systems that can be described by
a specific thermodynamic metric. For instance, a pseudo-Euclidean
structure of the form

\begin{equation}
G=\Theta^{2}+(\delta_{ab}E^{a}I^{b})(\eta_{cd}dE^{c}dI^{d})\label{eq:Gmetric}
\end{equation}
 with $\eta_{cd}=\mbox{diag}\left(-1,1,1,...,1\right)$, is Legendre
invariant (because of the invariance of the Gibbs 1-form) and induces
on ${\cal E}$ the metric 
\begin{equation}
g=\left(E^{f}\frac{\partial\Phi}{\partial E^{f}}\right)\left(\eta_{ab}\delta^{bc}\frac{\partial^{2}\Phi}{\partial E^{c}\partial E^{d}}dE^{a}dE^{d}\right)\label{eq:metricg}
\end{equation}

that describes systems characterized with second order phase transitions.
If the structure is an Euclidean metric on the form

\begin{equation}
G=\Theta^{2}+(\delta_{ab}E^{a}I^{b})(\delta_{cd}dE^{c}dI^{d})\label{eq:Gmetric-1}
\end{equation}
it is also Legendre invariant and induces on ${\cal E}$ the metric
\begin{equation}
g=\left(E^{c}\frac{\partial\Phi}{\partial E^{c}}\right)\left(\frac{\partial^{2}\Phi}{\partial E^{a}\partial E^{d}}dE^{a}dE^{d}\right)\label{eq:metricg-1}
\end{equation}

that describes systems with first order phase transitions.

\section{The Schwarzschild-AdS Black Hole}

The Einstein action with cosmological constant $\Lambda$ term is
given by

\begin{equation}
\mathcal{A}=\frac{1}{16\pi}\int d^{4}x\sqrt{-g}\left(R-2\Lambda\right),
\end{equation}
and the solution representing a static black hole is given by the
Kerr-AdS solution 

\begin{eqnarray*}
ds^{2} & = & -f\left(r\right)dt^{2}+\frac{dr^{2}}{f\left(r\right)}+r^{2}d\Omega^{2}
\end{eqnarray*}
where 

\begin{equation}
f\left(r\right)=1-\frac{2M}{r}-\frac{\Lambda}{3}r^{2}
\end{equation}

and $d\Omega^{2}=d\theta^{2}+\sin^{2}\theta d\varphi^{2}$. The horizons
are given by the condition

\begin{equation}
f\left(r_{H}\right)=1-\frac{2M}{r_{H}}-\frac{\Lambda}{3}r_{H}^{2}=0
\end{equation}

or

\begin{equation}
\frac{\Lambda}{3}r_{H}^{3}+2M-r_{H}=0\label{eq:horizon}
\end{equation}

In particular, for $\Lambda<0$, the largest positive root located
at $r=r_{+}$ defines the event horizon with an area

\begin{eqnarray}
A & = & 4\pi r_{+}^{2}.\label{eq:area}
\end{eqnarray}

The Smarr formula for the Schwarzschild-AdS black hole gives the relation

\begin{equation}
M=\sqrt{\frac{S}{4\pi}}\left(1-\frac{\Lambda}{3}\frac{S}{\pi}\right)\label{eq:fundamentalM}
\end{equation}

that corresponds to the fundamental thermodynamical equation $M=M\left(S,\Lambda\right)$
that relates the total mass $M$ of the black hole with the extensive
variables entropy $S=\frac{A}{4}$ and cosmological constant $\Lambda$
and from which all the thermodynamical information can be derived. 

In the geometric formulation of thermodynamics we will choose $E^{a}=\left\{ S,\Lambda\right\} $
and the corresponding intensive variables as $I^{a}=\left\{ T,\Psi\right\} $,
where $T$ is the temperature and
$\Psi$ is the generalized variable conjugate to the state parameter
$\Lambda$. In this way, the coordinates that we will use in the 5-dimensional
thermodynamical space ${\cal T}$ are $Z^{A}=\left\{ M,S,\Lambda,T,\Psi\right\} $.
The contact structure of ${\cal T}$ is generated by the 1-form

\begin{equation}
\Theta=dM-TdS-\Psi d\Lambda.
\end{equation}

No we will apply the GTD formalism to the Schwarzschild-AdS black hole using  $M$ as the thermodynamical potential.

\subsection{Mass Representation}

To obtain the induced metric in the space of equilibrium states ${\cal E}$
we introduce the smooth mapping

\begin{equation}
\varphi:\{S,\Lambda\}\longmapsto\left\{ M(S,\Lambda),S\Lambda,T\left(S,\Lambda\right),\Psi\left(S,\Lambda\right)\right\} 
\end{equation}

along with the condition $\varphi^{*}(\Theta)=0$ , that corresponds
to the first law $dM=TdS+\Psi d\Lambda$. This condition also gives
the relation between the different variables with the use of the fundamental
relation (\ref{eq:fundamentalM}). The Hawking temperature is evaluated
as

\begin{equation}
T=\frac{\partial M}{\partial S}=\frac{1}{4}\sqrt{\frac{1}{\pi S}}\left(1-\frac{\Lambda}{\pi}S\right),
\end{equation}

and the conjugate variable to $\Lambda$ is 

\begin{equation}
\Psi=\frac{\partial M}{\partial\Lambda}=-\frac{S}{3\pi}\sqrt{\frac{S}{4\pi}}.
\end{equation}

The variable $\Psi$ has dimensions of a volume, in fact using (\ref{eq:area})
we have $\Psi=-\frac{4}{3}r_{+}^{3}$ and it can be interpreted as
an effective volume excluded by the horizon, or alternatively a regularised
version of the difference in the total volume of space with and
without the black hole present \cite{entalpia1,entalpia2,entalpia3}.
Since the cosmological constant $\Lambda$ behaves
like a pressure and its conjugate variable as a volume, the term $\Psi d\Lambda$
has dimensions of energy and is the analogue of $VdP$ in the first
law. This suggests that after expanding the set of thermodynamic variables
to include the cosmological constant, the mass $M$ of the AdS black
hole should be interpreted as the enthalpy rather than as the total
energy of the spacetime. Therefore we are working the geometrothermodynamics
in the enthalpy representation.

\subsubsection{Second Order Phase Transitions }

Using (\ref{eq:Gmetric}),${\cal T}$ becomes a Riemannian manifold
by defining the metric 

\begin{equation}
G=\left(dM-TdS-\Psi d\Lambda\right)^{2}+\left(ST+\Psi\Lambda\right)\left(-dSdT+d\Lambda d\Psi\right).
\end{equation}

This metric has non-zero curvature and its determinant is $\det\left[G\right]=\frac{\left(ST+\Psi\Lambda\right)^{4}}{16}$.
To obtain the induced metric in the space of equilibrium states ${\cal E}$
we use equation (\ref{eq:metricg}), obtaining

\begin{equation}
g=\left(SM_{S}+\Lambda M_{\Lambda}\right)\left(\begin{array}{cc}
-M_{SS} & 0\\
0 & M_{\Lambda\Lambda}
\end{array}\right),\label{eq:g1}
\end{equation}
where subscripts represent partial derivative with respect to the
corresponding coordinate. However, note that the determinant of this
metric is null,
\begin{equation}
\det\left[g\right]=-M_{SS}M_{\Lambda\Lambda}\left(SM_{S}+\Lambda M_{\Lambda}\right)=0 ,\label{eq:detg}
\end{equation}
because $M_{\Lambda\Lambda}=0$. Therefore, this is not a suitable metric
for describing the thermodynamics of the black hole.

\subsubsection{First Order Phase Transitions }

Using (\ref{eq:Gmetric-1}),we define on ${\cal T}$ the euclidean
metric 

\begin{equation}
G=\left(dM-TdS-\Psi d\Lambda\right)^{2}+\left(ST+\Psi\Lambda\right)\left(dSdT+d\Lambda d\Psi\right).
\end{equation}

This metric has non-zero curvature and its determinant is again $\det\left[G\right]=\frac{\left(ST+\Psi\Lambda\right)^{4}}{16}$.

Equation (\ref{eq:metricg-1}) let us define the metric structure
on ${\cal E}$ as

\begin{equation}
g=\left(SM_{S}+\Lambda M_{\Lambda}\right)\left(\begin{array}{cc}
M_{SS} & M_{S\Lambda}\\
M_{S\Lambda} & M_{\Lambda\Lambda}
\end{array}\right).\label{eq:g2}
\end{equation}
Note that the determinant of this
metric is the non-null function
\begin{equation}
\det\left[g\right]=\left(M_{SS}M_{\Lambda\Lambda}-M_{S\Lambda}^{2}\right)\left(SM_{S}+\Lambda M_{\Lambda}\right)=-M_{S\Lambda}^{2}\left(SM_{S}+\Lambda M_{\Lambda}\right).\label{eq:detg-2}
\end{equation}

\subsection{Total Energy Representation}

Since $M$ is the enthalpy of the black hole and $Z^{A}=\left\{ \Phi,E^{a},I^{a}\right\} =\left\{ M,S,\Lambda,T,\Psi\right\} $,
$\left(a=1,2,3\right)$, one can obtain the total energy $U=U(M,S,\Phi)$ representation
by performing the partial Legendre transformation

\begin{equation}
Z^{A}\rightarrow\tilde{Z}^{A}=\left\{ \tilde{\Phi},\tilde{E}^{a},\tilde{I}^{a}\right\} 
\end{equation}
with

\begin{equation}
\begin{cases}
\Phi & =\tilde{\Phi}-\tilde{E}^{2}\tilde{I}^{2}\\
E^{2} & =-\tilde{I}^{2}\\
I^{2} & =\tilde{E}^{2},
\end{cases}
\end{equation}

that corresponds to the transformation $U=M-\Psi\Lambda$. The Gibbs
1-form becomes

\begin{equation}
\Theta=d\tilde{\Phi}-\delta_{ab}\tilde{I}^{a}d\tilde{E}^{b}=dU-TdS+\Lambda d\Psi
\end{equation}

from which the first law, 

\begin{equation}
dU=TdS-\Lambda d\Psi,
\end{equation}
is obtained when considering the space of equilibrium states ${\cal E}$.
In the total energy representation, the metric (\ref{eq:Gmetric})
becomes

\begin{equation}
G^{U}=\left(dU-TdS+\Lambda d\Psi\right)^{2}+\left(ST+\Psi\Lambda\right)\left(-dSdT+d\Lambda d\Psi\right).
\end{equation}

Equation (\ref{eq:metricg}) let us define a metric structure on ${\cal E}$
as

\begin{equation}
g^{U}=\left(SU_{S}+\Psi U_{\Psi}\right)\left(\begin{array}{cc}
-U_{SS} & 0\\
0 & U_{\Psi\Psi}
\end{array}\right),\label{eq:g3}
\end{equation}
where subscripts represent partial derivative with respect to the
corresponding coordinate.The determinant of this metric is 
\begin{equation}
\det\left[g^{U}\right]=-U_{SS}U_{\Psi\Psi}\left(SU_{S}+\Psi U_{\Psi}\right).\label{eq:detg-1}
\end{equation}

However, note that the total energy is

\begin{equation}
U=M-\Psi\Lambda=\sqrt{\frac{S}{4\pi}}
\end{equation}
and therefore $U_{\Psi\Psi}=0$. This imply $\det\left[g^{U}\right]=0$
and $g^{U}$ is not suitable as a thermodynamical metric.

\subsection{Entropy Representation}

In the context of GTD, it is also possible to consider the entropy
representation. In this case, the fundamental equation is given as
$S=S\left(M,\Lambda\right)$ and the Gibbs 1-form of the phase space
can be chosen as

\begin{equation}
\Theta^{S}=dS-\frac{1}{T}dM+\frac{\Psi}{T}d\Lambda.
\end{equation}

The space of equilibrium states ${\cal E}$ we introduce the smooth
mapping

\begin{equation}
\varphi^{S}:\{M,\Lambda\}\longmapsto\left\{ S(M,\Lambda),M,\Lambda,T\left(M,\Lambda\right),\Psi\left(M,\Lambda\right)\right\} 
\end{equation}

along with with the first law $\varphi^{S*}(\Theta^{S})=0$, which
gives the conditions

\begin{equation}
\frac{1}{T}=\frac{\partial S}{\partial M}
\end{equation}

and

\begin{equation}
\frac{\Psi}{T}=-\frac{\partial S}{\partial\Lambda}.
\end{equation}

Using (\ref{eq:Gmetric}),we define the metric 

\begin{equation}
G^{S}=\left(dS-\frac{1}{T}dM+\frac{\Psi}{T}d\Lambda\right)^{2}+\left(ST+\Psi\Lambda\right)\left(-dSdT+d\Lambda d\Psi\right).
\end{equation}

To obtain the induced metric in the space of equilibrium states ${\cal E}$
we use equation (\ref{eq:metricg}), that gives

\begin{equation}
g^{S}=\left(MS_{M}+\Lambda S_{\Lambda}\right)\left(\begin{array}{cc}
-S_{MM} & 0\\
0 & S_{\Lambda\Lambda}
\end{array}\right),\label{eq:gS}
\end{equation}
This time, the determinant of the metric is 
\begin{equation}
\det\left[g^{S}\right]=-S_{MM}S_{\Lambda\Lambda}\left(MS_{M}+\Lambda S_{\Lambda}\right).\label{eq:detg-3}
\end{equation}

\section{Phase Transitions and The Curvature Scalar}

Phase transitions are an interesting subject in the study of black
holes thermodynamics since there is no unanimity in their definition.
As is well known, ordinary thermodynamics defines phase transitions
by looking for singular points in the behavior of thermodynamical
variables. Following this argument, Davis \cite{key-3,davisthermo}
show that the divergences in the heat capacity indicate phase transitions.
For example, using equation (\ref{eq:fundamentalM}) we have that
the heat capacity for the Schwarzschild-AdS black hole is

\begin{equation}
C=T\frac{\partial S}{\partial T}=\frac{M_{S}}{M_{SS}}
\end{equation}

\begin{equation}
C=2S\frac{\left(\Lambda S-\pi\right)}{\left(\Lambda S+\pi\right)}.
\end{equation}

One can expect that phase transitions occur at the divergences of
$C$, i.e. at $M_{SS}=0$. For negative $\Lambda$ the divergence of $C$ corresponds to the well known Hawking-Page transition.
In geometrothermodynamics the apparition
of phase transitions is related with the divergences of the curvature
scalar $R$ in the space of equilibrium  states ${\cal E}$.
If we remember that $R$ always contains the determinant of the metric
$g$ in the denominator, we conclude that the zeros of $\det\left[g\right]$
could lead to curvature singularities (if those zeros are not canceled
by the zeros of the numerator). 

We have considered four different options for the metric in the case
of the Schwarzschild-AdS black hole. The first metric in the enthalpy
representation, (\ref{eq:g1}), has no inverse since its determinant
(\ref{eq:detg}) is always zero because $M\left(S,\Lambda\right)$
is linear in $\Lambda$. Therefore, this choice of metric is not good
to represent the thermodynamical system. The second choice of metric,
(\ref{eq:g2}), has the determinant given in equation (\ref{eq:detg-2}),
that is proportional to $M_{S\Lambda}$. However, this term is never
zero and the metric predicts no phase transitions. 

Using the total energy $U$ as thermodynamical potential we define
the metric given in (\ref{eq:g3}) but its determinant (\ref{eq:detg-1})
is null because $U\left(S,\Psi\right)$ does not depend explicitly
on $\Psi$.

Finally, we present the entropy representation in which $S=S\left(M,\Lambda\right)$
is the thermodynamical potential. This time the metric is chosen
as given by equation (\ref{eq:gS}) and its determinant, (\ref{eq:detg-3}),
is proportional to $S_{MM}$ and $S_{\Lambda\Lambda}$. This fact
makes clear the coincidence with the divergence of the heat capacity.
To see that, note that the heat capacity can be written in the entropy
representation as

\begin{equation}
C=T\frac{\partial S}{\partial T}=\frac{M_{S}}{M_{SS}}=-\frac{S_{M}^{2}}{S_{MM}},
\end{equation}
so the divergences of $C$ occur at $S_{MM}=0$. Even more, the curvature
scalar $R$ for the metric $g^{S}$ has the denominator

\begin{equation}
D=\left(MS_{M}+\Lambda S_{\Lambda}\right)^{3}S_{MM}^{2}S_{\Lambda\Lambda}^{2}
\end{equation}
which makes $R$ diverge when $S_{MM}=0$, corresponding to the Hawking-Page
phase transition \cite{key-2}.  Since we use a metric of the form given in (\ref{eq:metricg}) and following \cite{quevedo10}, we conclude that the divergence in the curvature scalar corresponds to a second order phase transition. 

Note that the factor $S_{\Lambda\Lambda}$
also appears in the denominator of the curvature scalar, but as can
be easily seen from (\ref{eq:fundamentalM}), this term gives

\begin{equation}
S_{\Lambda\Lambda}=\frac{\partial^{2}S}{\partial\Lambda^{2}}=\frac{2S^{3}}{9}\frac{\left(7\pi-5\Lambda S\right)}{\left(\pi-\Lambda S\right)^{3}}
\end{equation}
and therefore, for $\Lambda<0$ it never becomes zero. This fact shows that the consideration
of $\Lambda$ as a thermodynamical variable does not include new phase transitions in the system.

\section{Conclusion}

Quevedo's geometrothermodynamics is a differential geometric formalism
whose objective is to describe in an invariant manner the properties
of thermodynamic systems using geometric concepts. It indicates that
phase transitions would occur at those points where the thermodynamic
curvature $R$ is singular. However the curvature scalar depends on
the choice of the thermodynamical metric and as was shown in \cite{quevedo10}
the choices given in equations (\ref{eq:metricg}) and (\ref{eq:metricg-1})
apparently describe the second and first order phase transitions respectively.

In this work we applied the GTD formalism to the Schwarzschild-AdS
black hole, considering the cosmological constant as a new thermodynamical
state variable. In this approach, the mass of the black hole
is interpreted as its total enthalpy and when we apply the GTD formalism
we note that none of the chosen metrics describe phase transitions.
Performing a Legendre transformation to use the total energy as the
thermodynamical potential, we show that the curvature scalar does not  present
divergences. Finally, in the entropy representation we could
obtain a curvature scalar that diverges exactly at the point where
the Hawking-Page phase transition occurs. Since we use a metric of
the form given in (\ref{eq:metricg}) we conclude that this is a second
order phase transition. It is also important to note that the consideration
of $\Lambda$ as a thermodynamical variable does not include new phase
transitions in the system.

From the analysis above, it is clear that the phase manifold in the GTD formalism contains
information about thermodynamic systems; however, it is neccesary
a further exploration of the geometric properties in order to understand
where is encoded this information.\\

\emph{Acknowledgements}

This work was supported by the Universidad Nacional de Colombia. Hermes
Project Code 13038.

\end{document}